# Dissipative Kerr solitons in photonic molecules

Óskar B. Helgason, Francisco R. Arteaga-Sierra, Zhichao Ye, Krishna Twayana, Peter A. Andrekson, Magnus Karlsson, Jochen Schröder and Victor Torres-Company*

Department of Microtechnology and Nanoscience, Chalmers University of Technology, 41296 Gothenburg, Sweden

* Correspondence to: torresv@chalmers.se

**Many physical systems display quantized energy states. In optics, interacting resonant cavities show a transmission spectrum with split eigenfrequencies, similar to the split energy levels that result from interacting states in bonded multi-atomic, i.e. molecular, systems. Here, we study the nonlinear dynamics of photonic diatomic molecules in linearly coupled microresonators and demonstrate that the system supports the formation of self-enforcing solitary waves when a laser is tuned across a split energy level. The output corresponds to a frequency comb (microcomb) whose characteristics in terms of power spectral distribution are unattainable in single-mode (atomic) systems. Photonic molecule microcombs are coherent, reproducible, and reach high conversion efficiency and spectral flatness whilst operated with a laser power of a few milliwatts. These properties can favor the heterogeneous integration of microcombs with semiconductor laser technology and facilitate applications in optical communications, spectroscopy and astronomy.**

Optical microresonators are microscale cavities that confine light by resonant recirculation. They can be arranged in coupled arrays, with the dimensions of the unit cell precisely engineered by means of lithography. These coupled microresonator arrays exhibit characteristics that are analogous to the electron dynamics in strongly correlated many-body systems, allowing to study exotic states of matter whose Hamiltonians are difficult to access in their natural environment [1-4].

High-Q, small-volume optical microresonators lead to a large buildup of the intra-cavity intensity, easing the observation of optical nonlinear phenomena with low-power continuous-wave lasers [5,6]. A canonical example is the formation of dissipative Kerr solitons (DKS) [7] – self-enforcing solitary waves in nonlinear dissipative systems subject to an external energy supply [8]. DKS exist in cavities with normal as well as anomalous dispersion, and exhibit a rich diversity of soliton dynamics [10-15]. In the time-domain, DKS correspond to a localized waveform circulating at a rate commensurate to the cavity free-spectral-range (FSR) [9]. From a practical perspective, DKS in microresonators provide the means to realize microcombs [8] and a chip-scale version of the bi-directional coherent link between optics and microwave references that is the cornerstone of contemporary frequency metrology [16].

Most investigations of optical DKS have been carried out in single cavities designed with anomalous dispersion, where fundamental nonlinear physics [17] precludes attaining microcombs with an efficient distribution of the power available from the laser pump. In the normal dispersion regime, DKS exist in the form of interlocked switching waves that connect the homogeneous steady-state solutions of the nonlinear bistability diagram [11,18]. These waveforms feature high power-conversion efficiency [19] but their accessibility is limited to a narrow existence range of design and control parameters that are difficult to satisfy in single-mode cavities. These limits are intrinsic to single-mode dynamics and irrespective of the Q factor or nonlinear coefficient of the cavity. Here, we investigate DKS dynamics in a set of two linearly coupled high-Q silicon nitride microresonators (Fig. 1). The microresonators have identical cross-section geometry but different FSRs. The dispersion of the fundamental mode in the cavities is normal, resulting in a decreasing FSR with optical frequency. In absence of coupling, the resonant frequencies of the cavities, $\omega_\mu^{(a)}$ and $\omega_\mu^{(b)}$, are $\omega_\mu^{(a,b)} = \omega_0^{(a,b)} + \mu D_1^{(a,b)} + \mu^2 D_2^{(a,b)}/2 + \cdots$, with $D_1^{(a,b)}/(2\pi)$ representing the FSR of the cavity $(a,b)$, $D_2^{(a,b)}/(2\pi)$ the dispersion parameter that accounts from the deviation from a grid of evenly spaced frequencies, and $\mu$ is an integer number. Frequencies that satisfy $\omega_m^{(a)} = \omega_n^{(b)}$ are degenerate so that a laser tuned to one of them would be resonant in both cavities. By placing the microresonators in close proximity, the evanescent field of one cavity serves as a perturbation to the field in the adjacent one, resulting in linear coupling. The coupling lifts the frequency degeneracy and manifests in the emergence of a doublet of two new eigenfrequencies, corresponding to the eigenvalues of the supermodes, that are located at $\omega^{(s,as)} = [\omega_m^{(a)} + \omega_n^{(b)}]/2 \pm \left\{[\omega_m^{(a)} - \omega_n^{(b)}]^2/4 + |\kappa|^2\right\}^{1/2}$ with $\kappa$ being the coupling rate between microresonators. This arrangement of two microresonators placed in close proximity is often called "photonic molecule" in analogy to a diatomic multi-level system [20]. Our photonic molecule offers the flexibility of tuning the frequency offset $\omega_0^{(a)}$ by placing a heater on top of one of the microresonators. This in turn permits continuous tuning of the location of the degeneracy point $\omega_m^{(a)} = \omega_n^{(b)}$ and hence the frequency separation of the doublet $\omega^{(s)} - \omega^{(as)}$. The ability to lift the frequency degeneracy in linearly coupled nonlinear cavities has been used previously to impedance match the microcomb with an auxiliary drop port [21], as well as to facilitate phase matching with the pump frequency mode and initiate microcombs via modulation instability in the normal dispersion regime [22-24]. Our study (Fig. 2) reveals that the coupling changes dramatically the existence area of DKS.

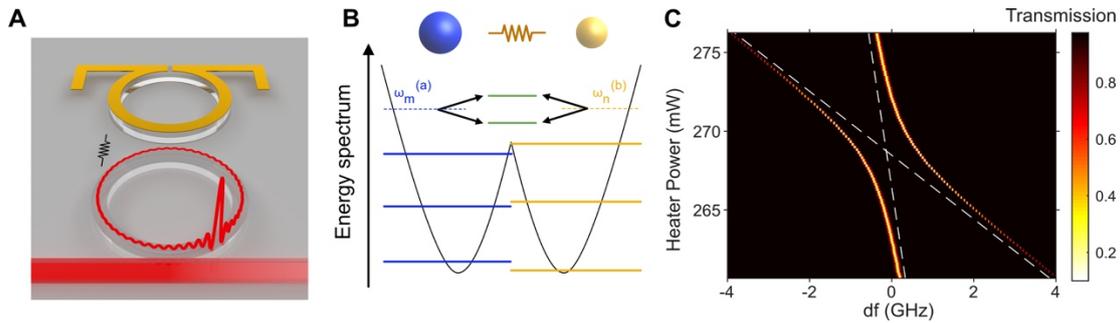

**Fig. 1. Photonic molecule based on linearly coupled microresonators.** (**A**) Schematic layout of the implementation, where coupling between microresonators is induced by placing them in close proximity. A heater on top of the auxiliary microresonator controls the frequency location of the degeneracy point. The coupling induces a controllable split resonance at this location, similar to a split energy level in a diatomic molecule (**B**). (**C**) Measurement of the frequency split as a function of heater power, illustrating the coupling between resonant frequencies. The auxiliary resonance shifts diagonally across the map, generating two supermodes as it gets more coupled to the main resonance. The main resonance shifts slightly with heater power due to thermal crosstalk.

DKS exist in the two-dimensional space given by the pump power and frequency detuning [25,26]. The existence range of DKS is rather narrow in normal-dispersion microresonators supporting a single transverse mode [26], as indicated by the area delimited by the solid black curve in Fig. 2. In a photonic molecule with fixed coupling rate ($\kappa/2\pi = 730$ MHz in this example; see other parameters in [27]), DKS exist in the region corresponding to the blue shade. A few important characteristics can be noted. First, this area is significantly broader than for the single-mode case, aiding the experimental accessibility of DKS. The simulations in Fig. 2 are conducted considering first-order dispersion walk-off between the two microresonators [27]. Photonic molecules enable the existence of asymmetric waveforms in both frequency and time domains that are not a steady-state solution of single-mode systems. This is because the dynamics of DKS in photonic molecules involves a complex interplay of solitary waves in both cavities. Second, the DKS dynamics in photonic molecules also leads to breathers and other forms of unstable waveforms. These regions of instability can be avoided if the level of the pump power is properly selected, allowing the transit into the stable DKS regime by simply tuning the laser at constant power into the doublet frequency starting from the blue side. This way, the DKS regime can be accessed in a deterministic manner at constant power after transiting through an initial stage of modulation instability, followed by the formation of Turing rolls. When a Turing roll is generated in the main cavity, another Turing pattern arises in the auxiliary, but with much weaker modulation. The transition to the DKS regime from the Turing rolls is accompanied by an abrupt decrease in intracavity power in the auxiliary cavity (see Fig. 3C), and the existence areas for these types of waveforms are clearly distinguished in the two-dimensional map. Finally, point 3 in the diagram illustrates the final states and compares the simulation (blue) with the measured microcomb spectrum (red). The steady-state time-domain waveform corresponds to a bright pulse with strong oscillations at the bottom, different to the sech profile encountered in anomalous dispersion cavities [7,9]. The high-power level is not a constant plateau either, as is common with dark-pulse Kerr combs [11]. The oscillations are a consistent feature among DKS in single-mode normal-dispersive microcombs, previously described as an interlock between the oscillatory tails of two switching waves, that keeps the waveform from collapsing [18]. The auxiliary cavity features another solitary wave synchronized to the pulse of the main cavity. Interestingly the kinks at the trailing edge of the pulse of the two waves in the cavities are in phase, but the leading edge appears exactly out of phase with the switching wave of the main cavity waveform. This is a consistent feature in these photonic molecule DKS as the time offset does not change with the detuning. The measured repetition rate for the waveform in point 3 is illustrated in Fig. 2D and it points to a coherent and stable waveform. Our study indicates the DKS dynamics in photonic molecules is predictable and can be adequately reproduced after analyzing the parameters of the coupled cavities by absorption spectroscopy [27].

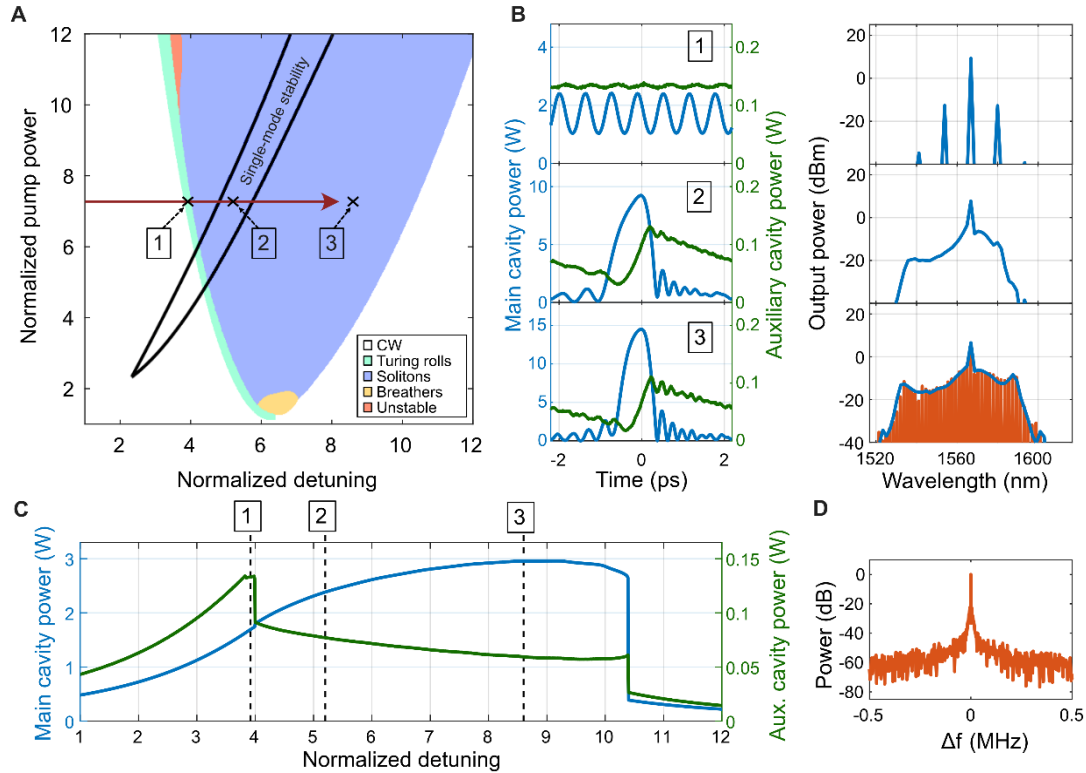

**Fig. 2. DKS dynamics in photonic molecules.** (**A**) Simulation of the stability chart of a photonic molecule in the normal dispersion regime with fixed coupling rate (730 MHz). At a fixed pump power between 2 and 10 in normalized units, the stable DKS region can be accessed in a deterministic fashion after transiting through a stable region of Turing rolls. The pump power and detuning are normalized as in [25]. (**B**) Simulated waveforms in points 1, 2 and 3 appear in the time domain (left column) and frequency domain (right column). The DKS waveforms display an asymmetric shape and the measured frequency spectrum (red) has clear sidebands. (**C**) The evolution of the average intracavity power for both main and auxiliary cavities as the laser is tuned into resonance. The locations of the microcombs of (**B**) are depicted by the dashed lines. (**D**) The beat note of the line spacing for the measured microcomb in point 3, measured with a resolution bandwidth of 2.5 kHz.

Figure 3 shows the photograph of one of the devices fabricated using a processing method for silicon nitride [28]. With this technique we can fabricate dispersion-engineered microring resonators with high optical-field confinement and high Q values. The fabrication process results in reproducible photonic molecules, as illustrated by the statistical analysis of the intrinsic Q presented in Fig. 3B for multiple resonances analyzed in the 1.5 μm band for 3 devices with identical geometry. The dispersion of the fundamental transverse mode in the main cavity mode is normal, as illustrated in Fig. 3C. This figure also shows two avoided mode crossings at 1535 nm and 1585 nm due to the controllable coupling between microresonators. In addition, there is an avoided mode crossing at 1572 nm owing to coupling between the fundamental transverse mode to a higher-order mode in the same microresonator, but this has only a minor influence on the results. The ability to select the frequency location of the split resonance results in broadly tunable DKS microcombs by frequency matching the pump laser to the location of the split energy level of the photonic molecule. This is shown in Fig. 3C, where we tune the photonic molecule microcombs over 40 nm within the telecommunications C band, from 1530 to 1570 nm, while maintaining an almost constant pump power (13 mW at the short wavelength and 18 mW near 1570 nm). The main cavity is over-coupled ($Q_{ex}$ = 2-4 million), resulting in a situation closer to effective critical coupling (loss rate equals coupling rate) when the DKS is generated.

Next, we investigate the power conversion efficiency of photonic molecule microcombs. DKS in normal dispersion microresonators have the ability to transform a larger fraction of the pump laser's power into useful comb power [19], a metric that is quantified by the conversion efficiency (defined as the ratio between on-chip output power excluding the pump divided by the input pump power). Over the frequency span displayed in Fig. 3D, the conversion efficiency increases from 32% to 49%. The increase in conversion efficiency is consistent with our simulations [27], and it is mainly caused by the increase in coupling rate at long wavelengths. Reaching such a high conversion efficiency is beneficial in several application of microcombs, but it can be misleading, since it is often dominated by the comb power concentrated in just a few lines, resulting in microcomb spectra with a non-smooth envelope. Having a uniform power distribution is important when deploying microcombs for coherent optical communications [28,29], the calibration of astronomical spectrographs [30] or dual-comb spectroscopy [31]. The results in Fig. 3 indicate that microcombs generated in photonic molecules result in an improved balance between conversion efficiency and spectral distribution compared to other state of the art microcombs [27]. A similar combination of conversion efficiency and spectral distribution can be found in the designs in Fig. 2 as well. Here, we take advantage of the increased existence regime at lower pump power, generating the comb from 2.5 mW on-chip power, see Fig. 4. This power level is compatible with recent advances in heterogeneous integration of III-V lasers on silicon nitride [32]. The conversion efficiency, $\eta$, is 36% resulting in a power per line > 12 µW, with little variation across the spectrum.

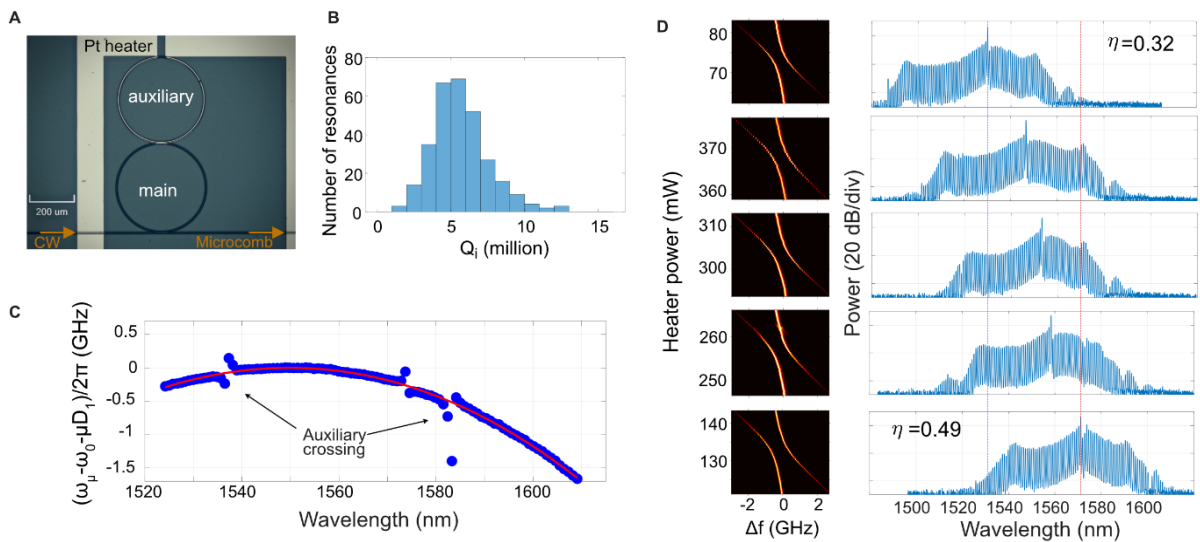

**Fig. 3. Control of photonic molecules and tunable microcombs.** (**A**) Photograph of fabricated device. The rings exhibit an FSR of 104.84 GHz and 106.9 GHz for the main cavity and auxiliary cavity, respectively. (**B**) Statistical analysis of the intrinsic Q of 3 nominally identical devices. The mean intrinsic value is 5.7 million. (**C**) Integrated dispersion of the main cavity mode, illustrating two split resonances (marked as auxiliary crossing points) that result from the coupling to the auxiliary cavity. (**D**) The frequency location of the crossing points can be selected with the heater power, and by tuning the pump laser to this location, wavelength tunable microcombs are generated.

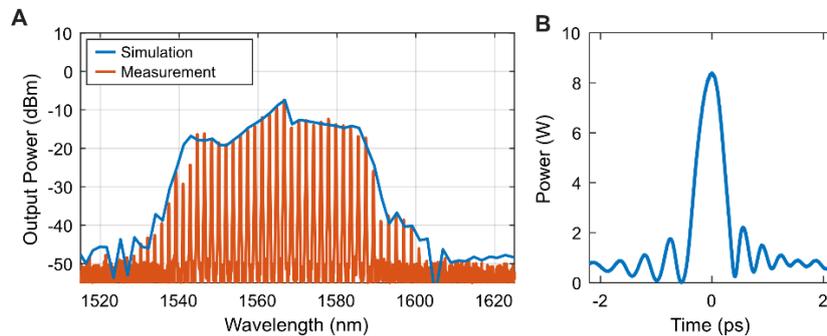

**Fig. 4. Flat photonic molecule microcombs.** (**A**) Measured on-chip power spectrum and (**B**) simulated time domain intensity profile inside the cavity. The comb is initiated by tuning the laser into main resonance from the blue side, while the auxiliary mode is blue detuned by 3 GHz from the main resonance.

In conclusion, we have demonstrated the formation of dissipative Kerr solitons in photonic diatomic molecules implemented in linearly coupled microresonators. The coupling rate between microresonators gives an additional degree of freedom for the synthesis of soliton states, expanding their existence range in the pump power–detuning bidimensional space with respect to a single mode (atomic) system. We demonstrate access to soliton states with high conversion efficiency and uniform power distribution, whose dynamics can be accurately predicted using the linear parameters of the cold molecular ensemble. Generalization to more complex molecules is possible, and advances in inverse nonlinear design in nanophotonics [33] offer the potential to effectively utilize the vast multi-dimensional parameter space for the synthesis of DKS microcombs with enhanced power distribution. The low power operation makes photonic molecule microcombs well suited for heterogeneous integration with semiconductor lasers, and their power efficiency and uniform distribution can be useful for applications such as optical communications.

**Funding**: European Research Council (ERC, CoG GA 771410); Swedish Research Council (2016-03960, 2016-06077); H2020 Marie Skodowska Curie Actions (Innovative Training Network Microcomb, GA 812818).

**Acknowledgements**: We are grateful to Prof. Andrew Weiner for the critical reading of the manuscript. The simulations for Fig. 2 were performed on resources at Chalmers Centre for Computational Science and Engineering (C3SE) provided by the Swedish National Infrastructure for Computing (SNIC).

# Dissipative Kerr solitons in photonic molecules – Supplementary Materials

Óskar Bjarki Helgason, Francisco R. Arteaga-Sierra, Zhichao Ye, Krishna Twayana, Peter A. Andrekson, Magnus Karlsson, Jochen Schröder and Victor Torres-Company

Department of Microtechnology and Nanoscience, Chalmers University of Technology, 41296 Gothenburg, Sweden

**S1 - Setup for dispersion measurement and analysis of cavity linewidth**

In order to determine the dispersion and Q-factors of the microcavities, we perform broadband measurements of the absorption spectra of the photonic molecules. The measurement setup we use is a modified version of [1] and it is displayed in Fig. S1. It involves three channels measured simultaneously in an oscilloscope. An external cavity diode laser (ECDL) is continuously swept in frequency, without mode-hops, from 1520 nm to 1620 nm where the resulting absorption profile of the device is recorded through a photodiode (PD1). Simultaneously, the ECDL beats against a commercially available self-referenced mode-locked fiber-laser with 250 MHz spacing. The interference between the ECDL laser and the self-referenced comb is measured in a balanced photodetector (BPD) and filtered through narrow bandpass radio-frequency filters (BP) centered at 30 MHz and 75 MHz, and finally recorded in the oscilloscope. This way, the oscilloscope provides calibration markers, allowing for a precise tracking of the ECDL sweep rate over time. The ECDL is also interfered with a continuous-wave laser locked to a known molecular absorption line near 1550.5156 nm. This beat note allows us to know the comb line number and calibrate the frequency axis.

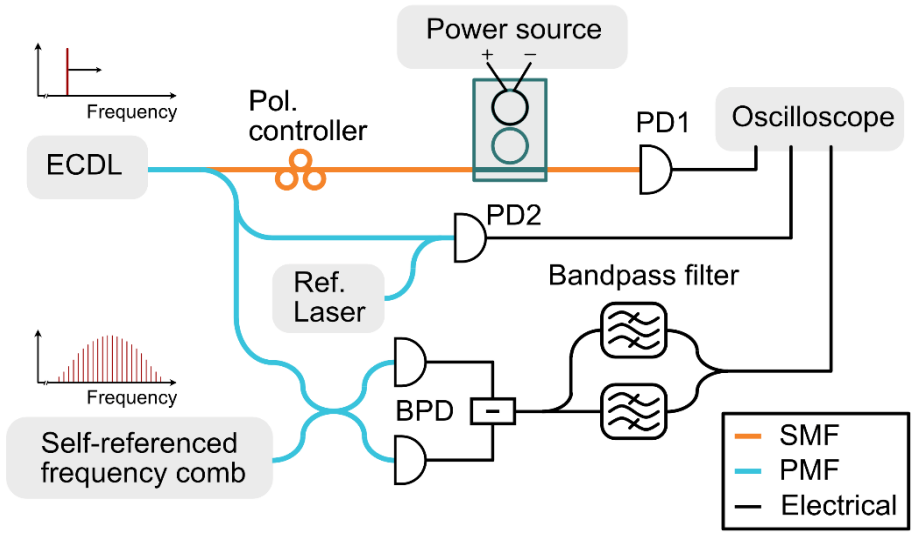

**Fig. S1. Setup for broadband absorption spectroscopy.** The setup consists of three different channels measured in an oscilloscope. One channel measures the absorption of the device as the frequency of the ECDL is swept. The second channel generates a beatnote between the ECDL and a reference laser, allowing for absolute wavelength calibration. In the third channel, the ECDL beats against the comb lines of a self-referenced mode-locked laser, enabling the precise monitoring of the sweep rate of the ECDL. SMF and PMF stands for single-mode fiber and polarization-maintaining fiber, respectively.

The results from the absorption spectra measurements are used to determine the quality factors and dispersion profile of the photonic molecules by characterizing the resonances. The position and linewidth of the majority of resonances can be fit well to a Lorentzian. The exceptions are a handful of resonances that display a resonance split due to parasitic linear coupling to a counter propagating mode. In those cases, we fit the resonances with a more accurate resonance doublet model [2]. Examples of resonance fitting are showed in Fig. S2. The quality factors of the main cavity are determined from the linewidths and extinction ratios of the resonance fits. The location of the main cavity resonances are used to fit a third-order dispersion profile using the equation $\omega_\mu = \omega_0 + \mu D_1 + \mu^2 D_2 /2 + \mu^3 D_3 /6$, with one such profile displayed in Fig. 3 in the main text. Here, $D_1 /(2\pi)$ describes the FSR, $D_2$ is related to the group velocity dispersion through $D_2 = -cD_1^2 \beta_2/n_g$ and $D_3$ to the third order dispersion through $D_3 = -cD_1^3 \beta_3/n_g + 3c^2 D_1^3 \beta_2/n_g^2$, where $n_g$ is the group index.

The parameters of the auxiliary cavity are more difficult to access compared to the main cavity, since the auxiliary resonances appear only in the spectral scans when they are coupled to resonances of the main cavity. Typically, only a handful of resonances appear, which allows a straightforward estimation of the auxiliary FSR. Other dispersion parameters and linewidths of the auxiliary cavity are not measured but are assumed to be close to that of the main cavity. This is a fair assumption since the two cavities share the same cross-sectional design.

The coupling rate between the two rings is determined from narrowband absorption spectroscopy, with essentially the same setup as displayed in Fig. S1, excluding the beatnote from the reference laser. The ECDL frequency is swept across a mode crossing over a span of a few GHz, recording the spectral characteristics for different heater powers. This results in a resonance split profile, one example being displayed in Fig. 1 in the main text, where the coupling rate is found as half the minimum separation between the two resonances.

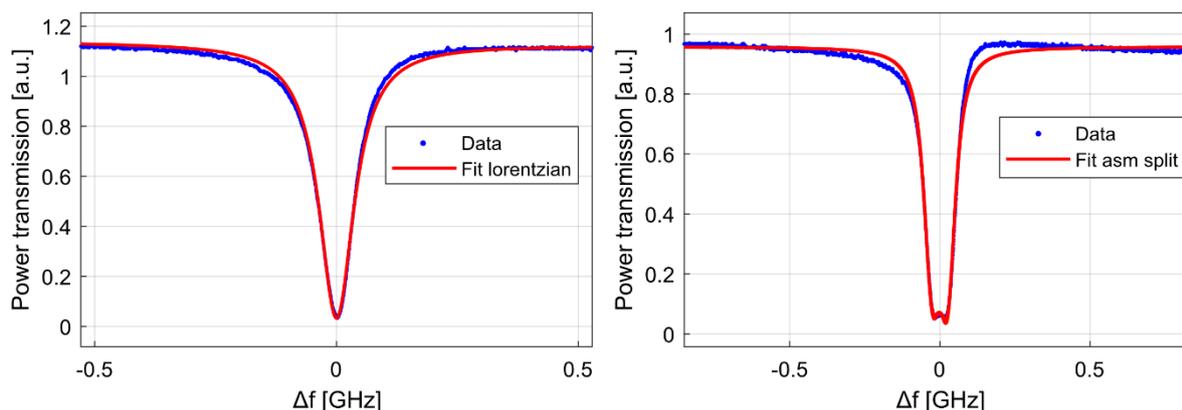

**Fig. S2. Fitting to resonances.** The figures show two models to fit resonances. On the left, a resonance is fitted with the Lorentz model. On the right, the data exhibits a resonance split induced by coupling to a counter propagating mode, which is fitted with a model for a resonance doublet.

### S2. Detailed cavity parameters

The two devices described in the main text of the paper were both fabricated in silicon nitride using a subtractive processing method, with a microheater placed on the auxiliary cavity. The waveguides of both cavities for both devices are designed with a height of 600 nm and 1650 nm width. Device 1 is presented in Fig. 2 and Fig. 4. It has a main ring with radius of 100 µm and auxiliary ring of radius=96 µm. The Device 2, presented in Fig. 3, has a main ring with radius=217 µm and an auxiliary ring of radius=213 µm. The dispersion and linewidth parameters of both devices were measured with the method described in part S1, with the results displayed in Table S1. The group velocity dispersion is $\beta_2 = 65\ ps^2/km$ for both devices. The intrinsic quality factor of Device 2 is based on measurements of three devices with identical design. The extrinsic quality factors are roughly estimated by averaging the seven $Q_{ex}$ values closest to the center wavelength. The table also shows the nonlinear Kerr parameter, $\gamma$, which is found using a mode-solver for the waveguide cross-section and materials. Additionally, we display the measured Q-factors and dispersion profile for both devices in Fig. S3. The figure clearly shows the wavelength dependence of the extrinsic quality factor.

Table S1. The parameters of the photonic molecules.

|  | $FSR_{main}$ (GHz) | $FSR_{aux}$ (GHz) | $Q_i$ (million) | $Q_{ex}$ (million) | $\kappa/2\pi$ (MHz) | $D_2/2\pi$ (MHz) | $D_3/2\pi$ (kHz) | $\gamma$ (Wm)$^{-1}$ |
|---|---|---|---|---|---|---|---|---|
| Device1 @ 1566.8 nm | 227.32 | 236.7 | 7.5 | 3.8 | 730 | -3 | 89 | 1.1 |
| Device2 @ 1550 nm | 104.84 | 106.9 | 5.7 | 3.1 | 485 | -0.6 | 3 | 1.1 |

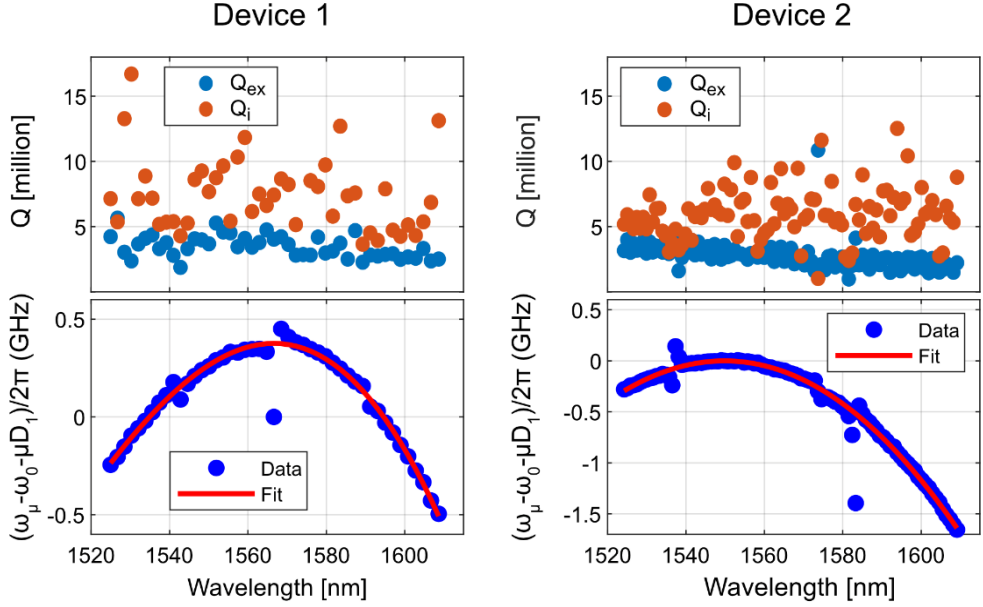

**Fig. S3. Measured Q factors and dispersion for device 1 and 2.** The upper figures show that the extrinsic quality factor decreases with wavelength. The lower figures show a third order fitting of the dispersion profile for both devices

### S3. Experimental setup for measuring the stability of the repetition rate

The repetition rate of the microcomb displayed in Fig. 3 is measured by electro-optic downconversion [3]. The setup for the measurement is displayed in Fig. S4. The microcomb is driven by the ECDL, then modulated with a phase modulator (PM) with frequency $f_{RF} = 25.146$ GHz, generating a sub-comb around each line of the microcomb. The radio frequency (RF) power supplied to the PM is sufficiently high such that the sub-combs overlap. The modulated microcomb is then filtered in an optical programmable filter (OPF), set to provide equally spaced band-pass filters, with a 0.3 nm passband and separation equal to the cavity FSR. Thus, only the desired overlapping lines of the sub-combs are measured, beating together in a high-speed photodiode before being recorded in a high-speed real-time oscilloscope. Finally, the beatnote is retrieved via simple Fourier analysis.

Fig. S5 shows a closer look at the downconverted beatnote recorded at two different instances. Each measurement spanned 0.4 ms which corresponds to a resolution bandwidth of 2.5 kHz. The plots show a clear beatnote, indicating stability on a short time scale. There is however a small shift in the location of the beatnote between the two recorded instances, which might be due to the pump shifting slightly in power or frequency over time. The repetition rate ($f_{rep}$) of the microcomb can be calculated from the beatnote through: $f_{rep} = 9f_{RF} + f_{beat} = 9 \cdot 25.146 + 1.004 = 227.318$ GHz.

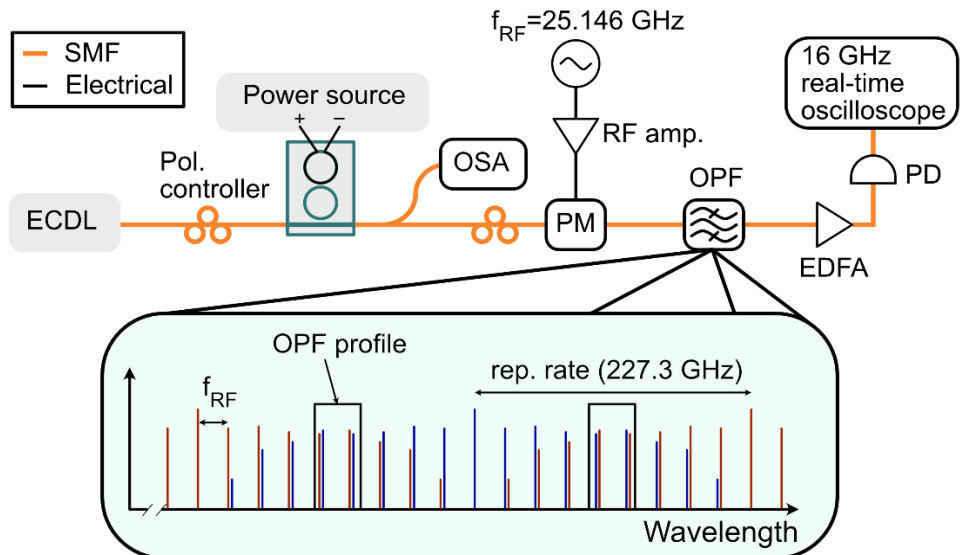

**Fig. S4. Setup for measuring the downconverted beatnote of a microcomb.** The ECDL pumps the coupled microcavities to generate a microcomb. The microcomb is phase-modulated at 25.146 GHz, resulting in overlapping sub-combs. Overlapping sections of multiple sub-combs are passed through the optical programmable filter (OPF). The sub combs beat together in a photodiode, which is then recorded in a high-speed real-time oscilloscope. The plot in the bottom of the figure shows the spectra of the phase-modulated microcomb as it enters the OPF, where adjacent sub-combs are represented with a different color for clarity. Only the lines within the OPF profile will pass through, where the separation between lines from the interleaved sub-combs create the downconverted beatnote.

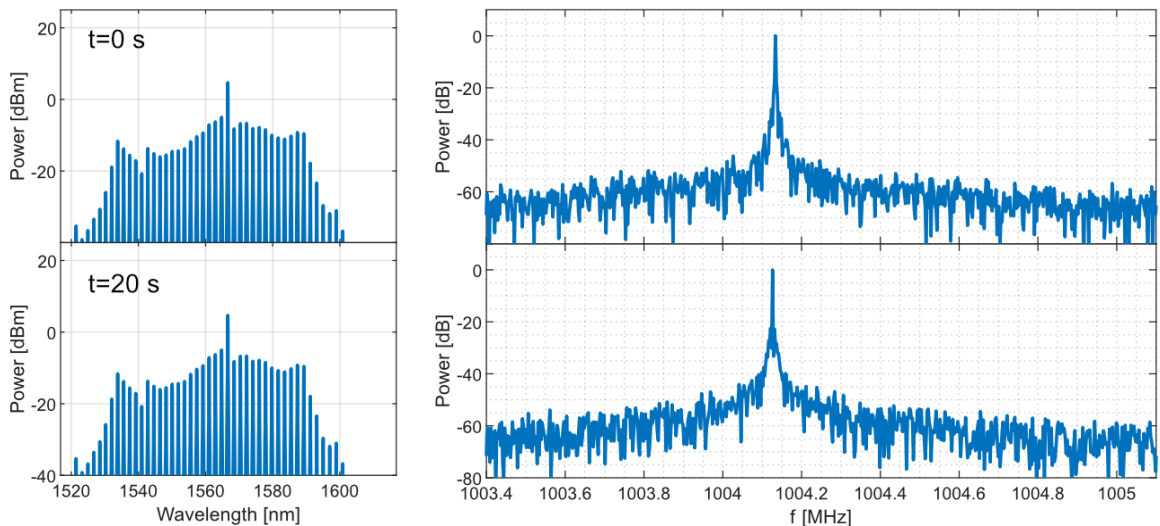

**Fig. S5. Measured downconverted repetition rate.** The upper figures show the comb state (left) and the downconverted repetition rate (right) at time instance t=0 s. The lower figures show a repeated measurement 20 s later. The spectra around the downconverted beatnote is clean, indicating low-noise operation.

## S4. Accessing DKS in photonic molecules

The DKS can generally be attained in the coupled microcavities by first tuning the heater on the auxiliary resonator such that the auxiliary resonance is slightly blue shifted from the main resonance (i.e. $\omega_0^{aux} > \omega_0^{main}$). The comb can then be generated deterministically by simply tuning the laser from the blue side into the main resonance. Once a comb state is achieved, the shape of the comb can be further optimized by tuning both the heater and the laser frequency.

Fig. S6 shows the path to initiating the microcomb displayed in Fig. 4. It shows the transmission of the coupled rings, detected by a photodiode and recorded in an oscilloscope. The oscilloscope recorded simultaneously the converted power, which measures a portion of the generated comb spectra, where the pump frequency has been filtered away using an optical programmable filter. The figure shows that the laser pushes first through the auxiliary resonance, which is blue detuned by roughly 3 GHz compared to the main resonance. The laser then approaches the main resonance, where the converted power rises when a DKS is generated.

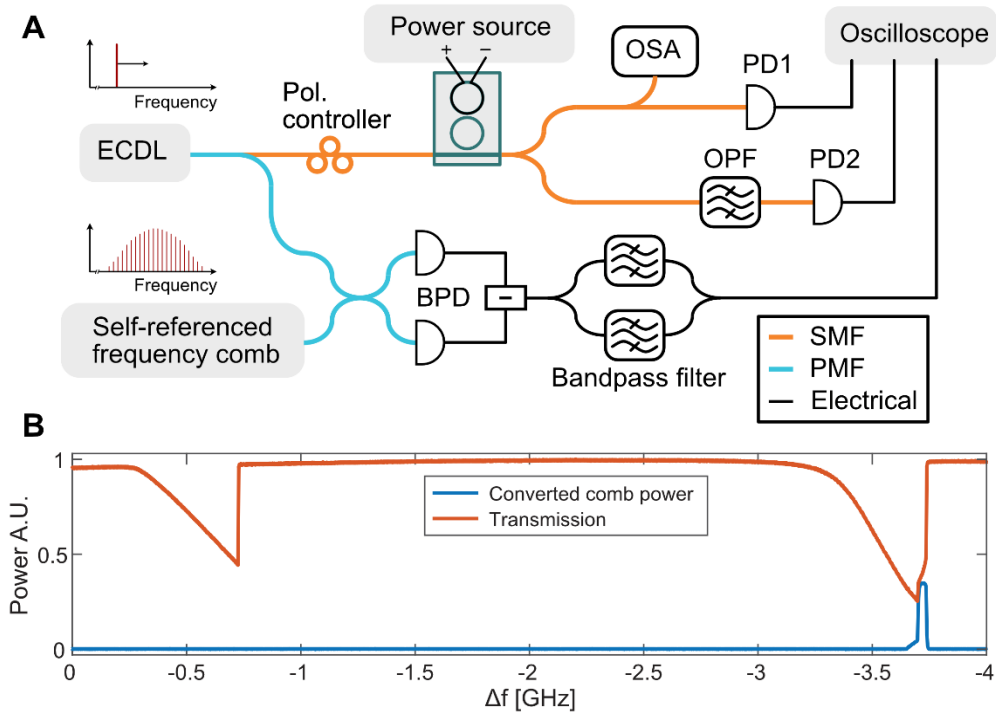

**Fig. S6. Initiation of a DKS comb state.** The diagram in **A** shows the setup used to acquire the traces of converted comb power and transmission. The transmission is measured in PD1, the converted power is measured in PD2 after filtering out the pump frequency using the OPF. The ECDL is swept in frequency via piezo control, with the change in frequency recorded using the interference between the ECDL and a self-referenced mode-locked laser. **B** shows the traces of converted comb power and transmission of the coupled cavities as the laser is tuned into the doublet of split resonances from the blue side. The dip in transmission corresponds to the laser pushing into resonance, where the resonance on the left (right) corresponds mainly to power building up in the auxiliary cavity (main cavity). The jump in converted power shows where a comb state is generated.

### S5. Numerical modeling of photonic molecule microcombs

A numerical verification of our comb states was conducted using the Ikeda map, expanded to include the auxiliary ring. The model is based on [5], involving two coupled rings (see Fig. S7), each having the same circumference ($L$), with the FSR difference between cavities emulated by using a different group index. The two coupling regions (bus waveguide-to-ring and ring-to-ring) are separated by distance of $L/2$. Thus, a full roundtrip of simulation involves (1) coupling between bus waveguide and main ring, (2) nonlinear propagation in both rings over a distance $L/2$, (3) coupling between main ring and auxiliary ring and (4) nonlinear propagation in both rings over a distance $L/2$.

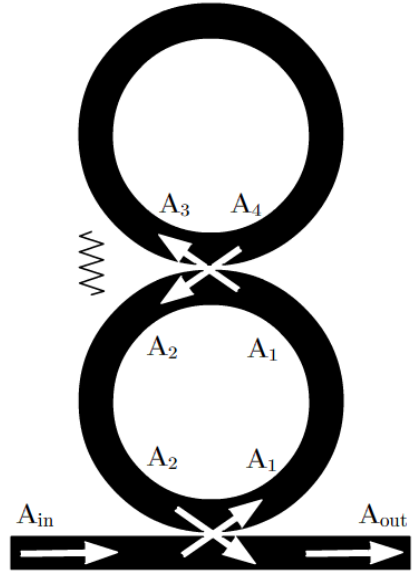

**Fig. S7.** Schematic of linearly coupled microresonators.

The bus waveguide-to-ring and ring-to-ring transmissions are calculated through coupled mode theory by using the corresponding coupling matrices expressed as

$$\begin{bmatrix} A_{out} \\ A_1 \end{bmatrix} = \begin{bmatrix} \sqrt{1-\theta_1} & i\sqrt{\theta_1} \\ i\sqrt{\theta_1} & \sqrt{1-\theta_1} \end{bmatrix} \begin{bmatrix} A_{in} \\ A_2 \end{bmatrix},$$

and

$$\begin{bmatrix} A_3 \\ A_2 \end{bmatrix} = \begin{bmatrix} \sqrt{1-\theta_2} & i\sqrt{\theta_2} \\ i\sqrt{\theta_2} & \sqrt{1-\theta_2} \end{bmatrix} \begin{bmatrix} A_1 \\ A_4 \end{bmatrix},$$

respectively, where $\theta_1 \approx (\frac{2\pi f_0}{FSR \cdot Q_{ex}})$ and $\theta_2 \approx (\frac{\kappa}{FSR})^2$ describe the portion of power coupled between waveguides, with $f_0$ being the pump frequency. The nonlinear propagation in each microresonator is described by the nonlinear Schrödinger equation (NLSE),

$$(\frac{\partial}{\partial z} + \frac{\alpha^{(a,b)}}{2} + i\delta^{(a,b)} + d^{(a,b)}\frac{\partial}{\partial T} + i\frac{\beta_2^{(a,b)}}{2}\frac{\partial^2}{\partial T^2} - \frac{\beta_3^{(a,b)}}{6}\frac{\partial^3}{\partial T^3} - i\gamma^{(a,b)}|A^{(a,b)}|^2)A^{(a,b)} = 0,$$

where $\alpha^{(a,b)}$ is the propagation loss, $d^{(a,b)}$ describes the pulse walk-off between the two cavities using the main cavity as reference, $\beta_2^{(a,b)}$ is the group velocity dispersion and $\gamma^{(a,b)}$ is the nonlinear Kerr parameter. $\delta^{(a,b)}$ describes the accumulated phase per unit length with regards to the pump laser, which can be transformed into frequency detuning via $\Delta f^{(a,b)} = FSR^{(a,b)}\delta^{(a,b)}L/(2\pi)$. The superscript, $a$ and $b$, correspond to main cavity and auxiliary cavity, respectively. In each round trip, a CW pump along with quantum noise consisting of one photon per spectral bin with random phase is coupled to the ring. The nonlinear propagation is solved numerically using the Runge-Kutta method in the interaction picture [4].

The model was used to simulate the DKS comb states in Fig. 2. The parameters used in these simulations were based on measurements of device 1, with the auxiliary resonance blue detuned 2.5 GHz from main resonance ($\Delta f^b - \Delta f^a = 2.5$ GHz) and $\kappa/2\pi = 730$ MHz, with parameters detailed in Tables S1 and S2. The intrinsic Q used in the simulations is slightly lower (6.6 million) than the measured mean intrinsic Q, but within experimental uncertainty (see Fig. S3). The laser detuning is estimated only qualitatively, by comparing the simulated and measured power transmission/conversion traces (see Fig. S8). A quantitative measurement of the detuning was not conducted due to challenges with distinguishing between thermal heating and Kerr nonlinearities.

The DKS comb simulated in Fig. 4 was also based on device 1, using the parameters found in Table S2. The pump power was set to 2.5 mW with the auxiliary resonance detuned 3 GHz from the main resonance. The comb is initiated by tuning into the main resonance from the blue side. The laser stops when it is red detuned by $\Delta f^a = 231.5$ MHz from main resonance and $\Delta f^b = 3.2315$ GHz from auxiliary resonance.

**Table S2. Parameters for numerical simulations of device 1**

|  | FSR (GHz) | Radius (μm) | $n_g$ |
|---|---|---|---|
| Main cavity | 227.33 | 100 | 2.099 |
| Aux. cavity | 236.84 | 100 | 2.015 |

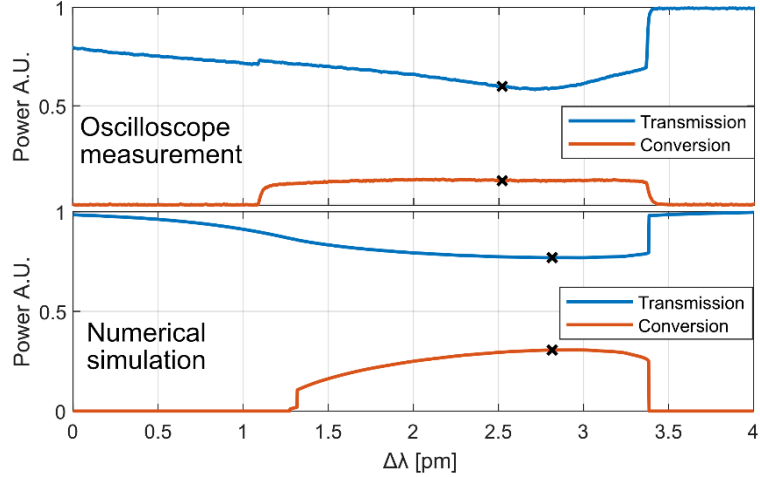

**Fig. S8. Transmission and converted power with regards to the tuned wavelength of the pump laser for both numerical simulation and oscilloscope measurement.** Marked on the traces are the estimated detuning locations of the microcomb number 3 displayed in Fig. 2B. Note that the measurement is expected to be different from numerical simulations since thermal heating is not captured in the simulation.

### S6. Analysis of wavelength dependence in conversion efficiency

The microcombs demonstrated in Fig. 3 (generated in device 2) displayed conversion efficiency that rises with wavelength. Using the numerical model introduced in S5, we replicate the shape and conversion efficiency of device 2 at two wavelengths, 1530 nm and 1566 nm. The simulations take into account the wavelength dependence of the pump power ($P_{in}$), extrinsic quality factor ($Q_{ex}$), coupling rate between rings ($\kappa$), and group velocity dispersion ($\beta_2$). The results are displayed in Fig. S9. The simulations match both the measured comb shape and conversion efficiency accurately, showing that the wavelength dependence of conversion efficiency can be predicted with our numerical model.

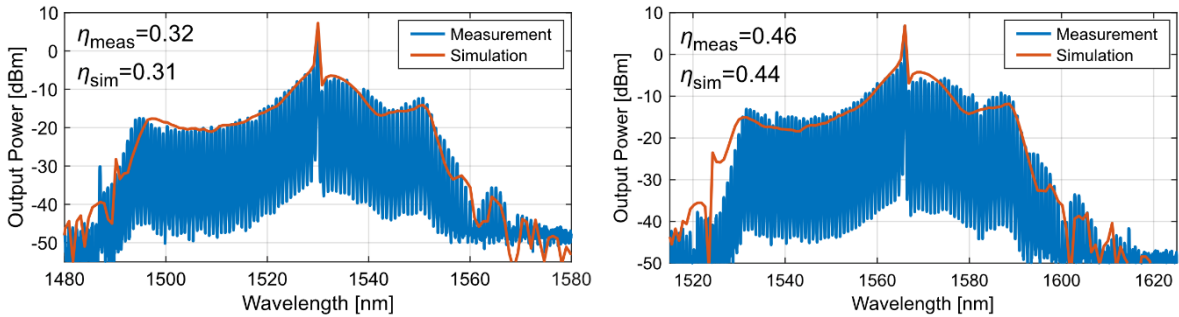

**Fig S9. Comparing numerical model with measurements of device 2.** The blue traces show the measurement results of microcombs generated in device 2 at 1530 nm (left) and 1566 nm (right).

### S7. Analysis of the stability chart

To generate the stability chart of Fig. 2A, we used the numerical model and parameters detailed in section S5. The simulations are carried out for a large set of coordinates in terms of normalized detuning and normalized power, defined as $F^2 = 8P_{in}\gamma L\theta_1/(\alpha L + \theta_1)^3$ and $\Delta = 2\beta_0^a L/(\alpha L + \theta_1)$, respectively. Here, $P_{in}$ is the input CW pump and $\alpha$ is the intrinsic loss of the main cavity. To ensure the detection of states spanning small regions, we calculated 568,000 points in the chart. Given the large number of simulations required, we used a computer cluster provided by Chalmers Centre for Computational Science and Engineering (C3SE).

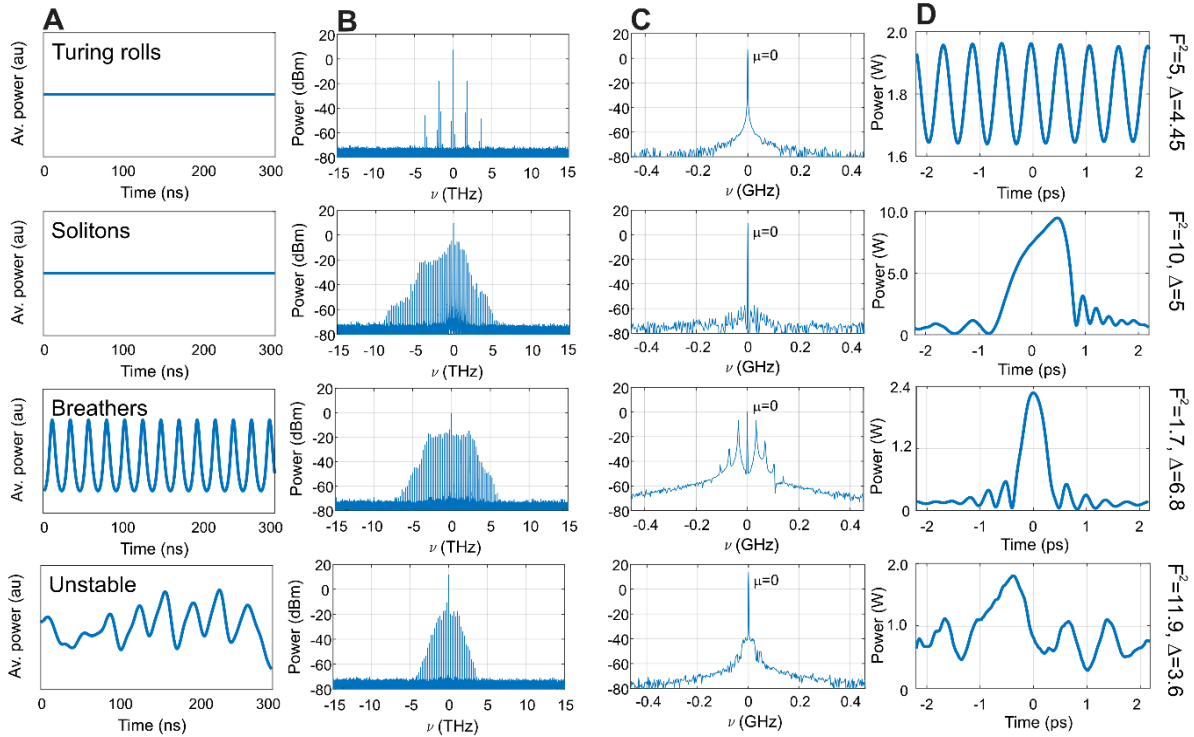

**Fig. S10. Features of each region shown in Fig 2A.** The four lines show examples from each region. The columns show different information for each sampling point: **(A)** average out-coupled power evolution from roundtrip-to-roundtrip for the last 300 ns, **(B)** out coupled power spectrum, **(C)** spectral components at µ=0 of the family of modes (zoom in of the out coupled optical power spectrum), and **(D)** intracavity intensity. All this information is used to characterize the steady-state Kerr combs of Fig 2.

For each power level, the simulation starts with normalized detuning of $\Delta = -1$, which is increased in a step-like manner up to $\Delta = 12$. In each detuning step, roundtrips equivalent to up to 800 ns are simulated to allow a possible step-response or change in comb-state to settle. Before taking the next detuning step, additional roundtrips are evaluated, equivalent to 300 ns in real-time, recording data that is later used to analyze the comb states. Examples of the recorded data are displayed in Fig. S10. It shows, from left to right, the average out-coupled power evolution from roundtrip-to-roundtrip, power spectrum, zoom-in of the spectral components around the central frequency ($\mu = 0$), and intracavity intensity obtained for four sample points of the Cartesian coordinates ($\Delta, F^2$). This information is used as input to an algorithm designed to characterize the state regions. The corresponding average out-coupled power evolution helps to establish the state as "stable" if the difference between the maximum and the minimum values is less than 0.1 percent of the arithmetic mean of elements of this evolution. Otherwise, it is "unstable". Using this definition, we identify solitons and Turing roll states as "stable", and breather states as a special case of "unstable" characterized by a periodic change of the out coupled power. The second stage distinguishes between Turing rolls or CS in the stable regime by analyzing the comb line spacing of the power spectrum, and breathers in the unstable regime by looking at the Fourier transform of the train of complex time-domain fields from roundtrip-to-roundtrip, and identifying pure or chaotic oscillations with the additional spectral components around $\mu = 0$ **(C)**. The final intensities and their evolutions can be used to corroborate our classification **(D)**.

## S8. Table comparing net conversion efficiencies in state of the art microcombs

In this section, we establish a comparative assessment of the conversion efficiency and spectral power distribution of our photonic molecule microcombs against state-of-the-art silicon nitride microcombs [6-8]. The conversion efficiency ($\eta$) describes how much of the pump power is converted to other comb lines. This figure of merit suffers the drawbacks that: 1) 100% efficiency corresponds to complete pump depletion; 2) does not account for uneven power distribution among lines. To account for these shortcomings, we supplement the information provided by this metric with the net conversion efficiency ($\eta_{net}$). The first step in calculating the net conversion efficiency is to find the comb line with the lowest power ($P_{min}$) within a desired bandwidth. Assuming all generated comb lines ($N$) were equalized to $P_{min}$, the net conversion efficiency is then calculated as the total power in the equalized spectra divided by the input power ($\eta_{net} = NP_{min}/P_{in}$). Note that this metric implicitly includes how the power is efficiently distributed among the comb lines, with 100% corresponding to a perfectly square comb generated in absence of loss.

The result of our comparison is shown in Table S3, where we analyzed the microcomb states presented in Fig. 4 and Fig. S9. The other literature examples correspond to another photonic molecule microcomb from normal dispersive rings [6], a normal dispersive DKS in the form of a dark pulse [7] and a bright DKS [8]. In each case, the number of lines is selected so that $\eta_{net}$ is close to maximum. Each comb has an optimum performance at different number of lines, which can skew the result, since the minimum line power drops with the number of lines $P_{min} \sim 1/N^2$ [9,10]. This means that the net conversion is expected to scale as $\eta_{net} \sim 1/N$. In order make a more fair comparison between the different examples, we normalize the net conversion efficiency to 100 lines according to $\eta_{net}^{100} = \eta_{net} \cdot N/100$ .

The results show up to an order of magnitude difference between the conversion efficiency and net conversion efficiency, especially for the normal dispersive microcombs, where the power is distributed less evenly among lines compared to the bright DKS. However, the bright DKS suffers from a low conversion efficiency to begin with, resulting in an even lower net conversion. The DKS generated in the photonic molecules in Fig. 4 and Fig. S9 outperform the other examples, showing a balanced combination of conversion efficiency and flatness.

Table S3. Comparing net conversion efficiencies from state of the art microcombs

|  | Photonic molecule | Photonic molecule | Photonic molecule | Dark pulse DKS | Bright DKS |
|---|---|---|---|---|---|
| Source | Comb in Fig. 4 | Comb in Fig. S8 | Kim et al. [6] | Fülöp et al. [7] | Liu et al. [8] |
| $\eta$ | 0.36 | 0.46 | 0.406 | 0.228 | 0.015 |
| $P_{in}$ (mW) | 2.5 | 17.8 | 188 | 115 | 50.4 |
| $N$ | 24 | 71 | 45 | 34 | 35 |
| $P_{min}$ (mW) | 0.0126 | 0.0198 | 0.07 | 0.0693 | 0.0115 |
| $\eta_{net}$ | 0.121 | 0.0789 | 0.0173 | 0.02 | 0.0066 |
| $\eta_{net}^{100}$ | 0.029 | 0.056 | 0.0078 | 0.0068 | 0.0023 |